# Statistical Timing Analysis for Latch-Controlled Circuits with Reduced Iterations and Graph Transformations

Bing Li, Ning Chen, Ulf Schlichtmann, *Member, IEEE*

*Abstract*—Level-sensitive latches are widely used in high-performance designs. For such circuits efficient statistical timing analysis algorithms are needed to take increasing process variations into account. But existing methods solving this problem are still computationally expensive and can only provide the yield at a given clock period. In this paper we propose a method combining reduced iterations and graph transformations. The reduced iterations extract setup time constraints and identify a subgraph for the following graph transformations handling the constraints from nonpositive loops. The combined algorithms are very efficient, more than 10 times faster than other existing methods, and result in a parametric minimum clock period, which together with the hold time constraints can be used to compute the yield at any given clock period very easily.

## I. INTRODUCTION

MANY Statistical Static Timing Analysis (SSTA) algorithms have been introduced in recent years to model process variations and analyze the circuit performance, resulting in a distribution providing delay-yield information.

Assuming that cell delays are linear functions of Gaussian random variables, first-order methods are proposed in [1]–[3]. This linear assumption simplifies arrival time computations at the expense of accuracy. Since the first-order approximation can only handle the first two moments of the random variables, the skewness of random variables is considered in [4] to improve the accuracy of modeling and arrival time propagation. For the same purpose the canonical linear form in [2] is extended in [5] to handle non-Gaussian parameters and nonlinear delay functions. Quadratic methods are also proposed in [6]–[9] where delay modeling and arrival time propagation use second-order polynomials. In [10] gate delays are modeled as linear functions of random variables identified by independent component analysis and the APEX method proposed in [11] is used to compute the PDF/CDF of the arrival times. Instead of computing the delay of the circuit directly, the method in [12] provides statistical bounds for the yield of the circuit. The method in [13] also uses statistical bounds to predict the yield even at the pre-placement design stage.

The statistical methods discussed up to now handle combinational circuits and therefore can be easily adapted for timing analysis of circuits based on edge-triggered flip-flops. In practice level-sensitive latches are also widely used in high-performance designs. Because signals can pass the latches transparently critical paths may span multiple stages. This makes the timing analysis for such circuits more complex, requiring not only the statistical maximum and sum computations but also new algorithms handling the latch transparency, because the statistical computations cause the algorithms for static timing analysis of such circuits, for example, the linear programming method in [14], to be infeasible.

Several methods have been proposed for statistical timing analysis of latch-controlled circuits. The method in [15] computes the bit-error rate across an interconnect pipelined by level-sensitive latches. In [16] arrival times are propagated across latches to verify the timing constraints. But both methods do not process feedback loops, which are very common in sequential designs. The method in [17] solves this timing analysis problem using constraint graphs. The yield of the circuit is computed by verifying that no positive or negative loops exist in the corresponding constraint graphs and a backward edge breaking algorithm is proposed to capture these constraints. In [18] a statistical Bellman-Ford algorithm is used to verify the loop constraints combining graph decomposition and limiting the number of traversal stages to reduce runtime. Another method [19] proposes to run the iterations of timing propagation until the iteration means, which are defined in [19] to measure the convergence of the iterations, are stable.

For large circuits, however, the methods above are still computational expensive because of the large number of loops in the constraint graph [17] or the iteration convergence constraint [19]. Additionally, these methods can only compute the timing yield of the circuit at a given clock period. To obtain the complete yield curve, these methods need to be run several times with different clock periods, therefore increasing the runtime further. Because only the yields at given clock periods can be known using these methods, they are not suitable for hierarchical statistical timing analysis [20], [21], where the timing constraints of submodules are extracted separately and the yield of the complete circuit is computed only from these extracted constraints together with the correlation between them. This module-to-module correlation is lost inevitably using the existing methods for latch-controlled circuits.

In this paper, we propose a novel method to analyze the

The authors are with the Institute for Electronic Design Automation, Technische Universität München, Arcisstr. 21, Munich 80333, Germany.

This research was partially supported by the German Research Foundation (DFG) as part of the Transregional Collaborative Research Centre "Invasive Computing" (SFB/TR 89).

timing of latch-controlled circuits. The main contributions of this paper are as follows: 1) The proposed method computes a parametric minimum clock period whose statistical properties, such as mean and variance, are directly available so that the yield of the circuit at any given clock period can easily be evaluated. Since the correlation information is preserved by the statistical variables describing the minimum clock period, for example, in the canonical linear form in [2], the proposed method can also be used in hierarchical design flows. 2) With reduced iterations and graph transformations operating on a subgraph identified after the iterations the proposed method captures timing constraints more than 10 times faster than other previously published methods while maintaining the accuracy and providing more details. 3) In the handling of the constraints from nonpositive loops no feedback loop affecting the minimum clock period is missed so that the possible accuracy degradation can be avoided, which may happen in some circuits where the loops imposing dominant timing constraints may not be captured in other methods.

The rest of this paper is organized as follows. In Section II we give an introduction to static timing analysis of latch-controlled circuits. The timing constraints described in this section are the same in static and statistical timing analysis and form the basis of the proposed method. In Section III we explain our method including clock modeling, timing constraint decomposition, arrival time iterations for setup time constraints, graph transformations for loop constraints, and the extraction of the conservative hold time constraints. We show experimental results in Section IV and conclude our method in Section V.

## II. TIMING CONSTRAINTS OF LATCH-CONTROLLED CIRCUITS

In this section, we provide an overview for timing analysis of latch-controlled circuits. We also explain the static timing analysis algorithm from [22] in detail. This algorithm will be used as the basis to construct the proposed method in Section III.

### A. Basic Timing Formulations

Common in sequential circuits, a latch-controlled circuit is composed of combinational logic separated by latches functioning as synchronizers. The timing structure of such a circuit can be represented by a *Reduced Timing Graph* (RTG) [19], where a node represents a latch and an edge represents either the maximum or minimum combinational delay between a pair of nodes for verifying setup time or hold time constraints, respectively. Fig. 1 shows an example of an RTG with four latches.

In the RTG, if a node $j$ has an edge to $i$, $j$ is called a *fanin node* of $i$, denoted by $j \to i$. A *loop* in the RTG is defined as a directed path starting from and ending at the same node, for example, $1 \to 3 \to 2 \to 1$ in Fig. 1. In this paper the mentioned loops are all simple loops, meaning that no subloop exists on the path. If there is only one node on a loop, the loop is called a *self loop*. For example, node 2 in Fig. 1 has a self loop.

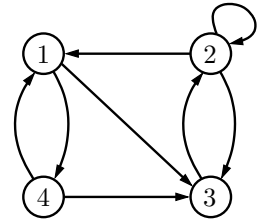

Fig. 1. Reduced timing graph

To evaluate the timing performance of a latch-controlled circuit, the complete timing constraints allowing multiphase clocks are specified in [14], namely the SMO model, which is thoroughly investigated in [22]. The mechanism of level-sensitive circuits does not change when process variations are considered except that all the delays and setup and hold times become statistical. The details of the SMO model are explained as follows.

In timing analysis of latch-controlled circuits, all arrival times are represented in the *local time zone* [23], [24]. The clock signals of two successive latches $j$ and $i$ and the corresponding local time zones are illustrated in Fig. 2. In this paper, we assume the active clock level of latches is '1' and the starting time of each local time zone is the time when the clock signal of the latch switches from '1' to '0'. When the clock is active, the latch is in transparency allowing signal propagation. Therefore the rising clock edge is called *enabling edge*, denoted by $r_i$ in the local time zone of $i$. At the falling clock edge the signal at the input is latched and thereafter no signal propagation is possible, so that this edge is called *latching edge*. According to the definition of local time zone, the time of the latching edge is always equal to the clock period $T$. $E_{j,i}$ is the *phase shift* of the two clocks and used to transform arrival times among different local time zones.

The parameters and constraints used in the SMO model [14] are listed in Table I, where some definitions and expressions are directly taken from [22] for consistency. In the SMO model, the time that a data signal starts to propagate to the next latch is called *departure time* and denoted by $D_i$ for the latest one and $d_i$ for the earliest one in Table I. Both values are defined in the local time zone of latch $i$. If the arrival time of the signal at latch $i$ is earlier than the enabling edge $r_i$, the data signal must wait until $r_i$ to start the propagation. Therefore, the departure time at latch $i$ is computed by (1) and (5) in Table I. Assume latch $j$ has at least one combinational path to $i$. The latest arrival time from $j$ to $i$ is computed by $D_j + \Delta_{j,i} - E_{j,i}$, where $\Delta_{j,i}$ is the maximum combinational delay from $j$ to $i$ and $-E_{j,i}$ converts the arrival time to the local time zone of $i$. $\Delta_{j,i} - E_{j,i}$ is denoted by $\Lambda_{j,i}$ in (2). When all fanin nodes of $i$ are considered, the latest arrival time at $i$

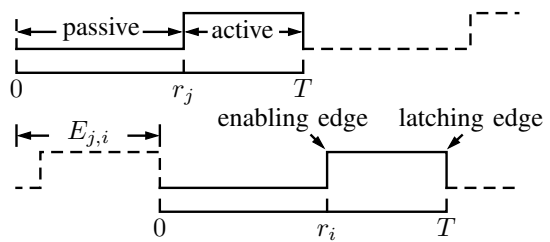

Fig. 2. Local time zone and clock phase shift

TABLE I
TIMING CONSTRAINTS OF LATCH-CONTROLLED CIRCUITS

**Variables**

| | |
|---|---|
| $n$ | number of latches in the circuit |
| $T$ | clock period |
| $r_i$ | time of the enabling clock edge in local time zone |
| $E_{j,i}$ | phase shift, time difference between starting times of clock phases $j$ and $i$ |
| $D_i$ | latest departure time from latch $i$ |
| $A_i$ | latest arrival time at latch $i$ |
| $\Delta_{j,i}$ | maximum combinational delay from latch $j$ to $i$ |
| $\Lambda_{j,i}$ | edge weight in the latest RTG |
| $s_i$ | setup time of latch $i$ |
| $d_i$ | earliest departure time from latch $i$ |
| $a_i$ | earliest arrival time at latch $i$ |
| $\delta_{j,i}$ | minimum combinational delay from latch $j$ to $i$ |
| $\lambda_{j,i}$ | edge weight in the earliest RTG |
| $h_i$ | hold time of latch $i$ |

**Setup Time Constraints**

$$D_i = \max\{A_i, r_i\} \quad (1)$$
$$\Lambda_{j,i} = \Delta_{j,i} - E_{j,i} \quad (2)$$
$$A_i = \max_{j \to i}\{D_j + \Lambda_{j,i}\} \quad (3)$$
$$A_i \leq T - s_i \quad (4)$$

**Hold Time Constraints**

$$d_i = \max\{a_i, r_i\} \quad (5)$$
$$\lambda_{j,i} = \delta_{j,i} - E_{j,i} \quad (6)$$
$$a_i = \min_{j \to i}\{d_j + \lambda_{j,i}\} \quad (7)$$
$$a_i \geq h_i \quad (8)$$

is computed by (3), where the maximum is performed over all fanin nodes of $i$. To guarantee the proper function of latches, the latest arrival time should be the setup time $s_i$ earlier than the latching clock edge. At latch $i$, this constraint is defined by (4). The definitions for hold time analysis can be explained similarly, except that there are both maximum and minimum computations in (5) and (7). This is the source of complexity for hold time analysis and will be explained in Section II-C. The arrival times are computed by adding $\Lambda_{j,i}$ and $\lambda_{j,i}$ to the departure times of the last latch stage as shown by (3) and (7). For convenience we assign $\Lambda_{j,i}$ and $\lambda_{j,i}$ as the *edge weight* of an edge in the RTG for the setup time or hold time verifications, respectively.

*B. Analyzing Setup Time Constraints*

According to (1) and (3) in Table I, the arrival times between successive latches are dependent. To find the minimum clock period of the circuit, the method in [14] uses linear programming to solve a relaxed formulation. But this method does not work when all the delays are random variables. In [22] an iterative method is proposed to verify the timing constraints with a given clock period for static timing analysis. We first show the details of this algorithm as listed in Algorithm 1 and use it later to explain our timing constraint decomposition in Section III.

At the beginning of this algorithm, the arrival times and departure times of all latches are initialized in L1–L4. After reset, data signals at all latches start to propagate from their enabling clock edges. The main iterations L5–L10 simulate the update of arrival times in the first $n$ clock cycles. In each iteration, the arrival times and departure times are computed from the ones in the last iteration, exactly the behavior of latch-controlled circuits.

At the $n$th iteration, Algorithm 1 checks if the setup time constraints are met based on the proof in [22, Lemma 2.2] that the arrival times are nondecreasing during the iterations, that is, $A_i^n \geq A_i^{n-1} \cdots \geq A_i^1$. Additionally, the algorithm checks if a positive loop exists in the RTG with the condition $D_i^n \neq D_i^{n-1}$ in L11–L15. If such a loop exists, the circuit can not work because setup time constraints will be violated after sufficient loop traversals. To explain this constraint further, we use the following definitions.

**Definition 1.** *For a path in the RTG with $n_l$ edges and $n_l + 1$ nodes $j_0, j_1, \ldots j_{n_l}$ the **cumulative weight** is defined as $\sum_{k=1}^{n_l} \Lambda_{j_{k-1}, j_k}$, that is, the sum of all the edge weights. If the path forms a loop so that $j_0 = j_{n_l}$ the path contains only $n_l$ nodes.*

**Definition 2.** *A **nonpositive loop** is a loop with nodes $j_0, j_1, \ldots j_{n_l}$, $j_0 = j_{n_l}$ and $\sum_{k=1}^{n_l} \Lambda_{j_{k-1}, j_k} \leq 0$, that is, the cumulative weight is nonpositive.*

According to L7 and L8 in Algorithm 1, the departure times at the nodes $j_0, j_1, \ldots j_{n_l}$, $j_0 = j_{n_l}$ on the loop meet

$$D_{j_k}^k \geq A_{j_k}^k \geq D_{j_{k-1}}^{k-1} + \Lambda_{j_{k-1}, j_k}, \quad k = 1, 2, \ldots n_l. \quad (9)$$

The departure time at $j_0$ after applying (9) across the loop for $k = 1, 2, \ldots n_l$ meets

$$D_{j_0}^{n_l} = D_{j_{n_l}}^{n_l} \geq D_{j_0}^0 + \sum_{k=1}^{n_l} \Lambda_{j_{k-1}, j_k}. \quad (10)$$

The sum in (10) is the cumulative weight of the loop. If this weight is positive, the arrival time at $j_0$ becomes large enough to violate the setup time constraints at the following latches after the loop is traversed sufficient times. Therefore, the cumulative weights of all the loops in the RTG must be

---

**Algorithm 1**: Iterative constraint verification

L1   **foreach** *node $i$ in RTG* **do**
L2       $A_i^0 = -\infty$
L3       $D_i^0 = r_i$
L4   **end**
L5   **for** $m = 1$ **to** $n$ **do**
L6       **foreach** *node $i$ in RTG* **do**
L7           $A_i^m = \max_{j \to i}\{D_j^{m-1} + \Lambda_{j,i}\}$
L8           $D_i^m = \max\{A_i^m, r_i\}$
L9       **end**
L10  **end**
L11  **foreach** *node $i$ in RTG* **do**
L12      **if** $D_i^n \neq D_i^{n-1}$ *or* $A_i^n > T - s_i$ **then**
L13         **return** *false*
L14      **end**
L15  **end**
L16  **return** *true*

verified being nonpositive to guarantee the proper behavior of the circuit. This verification is very challenging due to the number of loops in the RTG for large circuits.

In Algorithm 1 the nonpositivity of the loops is verified using the condition $D_i^n \neq D_i^{n-1}$ in L11–L15. If any node in the RTG meets this condition, there is at least one positive loop and some timing constraints will be violated. This verification can be explained as follows. If a positive loop exists in the RTG, there is at least a node $i$ with departure time $D_i^n > D_i^{n-1}$ [22], [25]. This can be proved using contradiction with an example adapted from [25]. Assume a positive loop having nodes $j_0, j_1, \ldots j_{n_l}$ with $j_0 = j_{n_l}$ and at all these nodes $D_{j_k}^n \leq D_{j_k}^{n-1}, 1 \leq k \leq n_l$. According to L7 and L8 of Algorithm 1 we also have

$$D_{j_k}^n \geq A_{j_k}^n \geq D_{j_{k-1}}^{n-1} + \Lambda_{j_{k-1},j_k}. \quad (11)$$

Adding both sides of (11) for all the nodes on the loop and with the fact that $j_0$ and $j_{n_l}$ are the same node, we have

$$\sum_{k=1}^{n_l} D_{j_k}^n \geq \sum_{k=1}^{n_l} D_{j_{k-1}}^{n-1} + \sum_{k=1}^{n_l} \Lambda_{j_{k-1},j_k} \iff \quad (12)$$

$$\sum_{k=1}^{n_l} (D_{j_k}^n - D_{j_k}^{n-1}) \geq \sum_{k=1}^{n_l} \Lambda_{j_{k-1},j_k}. \quad (13)$$

Because $D_{j_k}^n \leq D_{j_k}^{n-1}$ as assumed above, we can conclude that

$$0 \geq \sum_{k=1}^{n_l} \Lambda_{j_{k-1},j_k}. \quad (14)$$

This contradicts that the loop is positive so that there must be at least a node with $D_{j_k}^n > D_{j_k}^{n-1}$. From this reasoning, the condition for nonpositive loops is that every node meets $D_i^n \leq D_i^{n-1}$. Additionally, we have $D_i^m \geq D_i^{m-1}, 1 \leq m \leq n$ during the iterations according to [22], so that the nonpositive loop constraint becomes $D_i^n = D_i^{n-1}$.

Because the largest loop in the RTG contains no more than all the $n$ nodes in the RTG, the nonpositive loop constraints should be checked at the $n$th iteration to guarantee all loops are covered. If there is a violation at any node, Algorithm 1 returns $false$ asserting the existence of positive loops. If there is no positive loop in the RTG, the arrival times do not increase after the $n$th iteration because $D_i^n = D_i^{n-1}$ at any node, so that all timing constraints in the following iterations are guaranteed. Details of extracting setup time constraints will be given in Section III-B to III-E.

### C. Analyzing Hold Time Constraints

The hold time constraint at a latch $i$ requires that the earliest arrival time must be later than the hold time $h_i$ after the latching clock edge as specified by (8). Fig. 3 shows the concept of the hold time constraint from $j$ to $i$. After reset, the data signal starts to propagate from the rising clock edge $r_j$ and the arrival time at $i$ is updated in each following clock cycle. In [22] it is proved that the arrival time does not decrease during the updating, that is, $a_i^m \geq a_i^{m-1} \cdots \geq a_i^1$, and converges after a certain number of clock cycles. From this observation, two types of formulation for verifying hold time constraints have been proposed previously: the aggressive formulation and the conservative formulation.

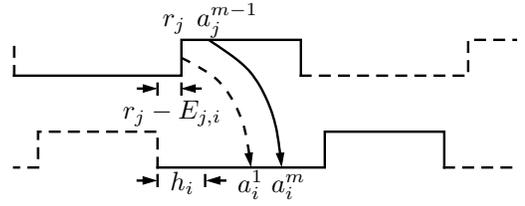

Fig. 3. Hold time constraints

The aggressive formulation verifies the converged arrival times with the hold time constraints. For example, if we assume the arrival times converge at iteration $m$, the constraint for latch $i$ is $a_i^m \geq h_i$. On the contrary, the conservative formulation verifies the constraint using $a_i^1 \geq h_i$ for latch $i$. The two formulations are illustrated in Fig. 3 where the dashed arrow shows the conservative one.

Because the earliest arrival times do not decrease during the iterations, verifying the circuit using the aggressive formulation leads to a higher yield, but with the risk that the hold time constraints at some latches may be violated during the first clock cycles when the arrival times have not converged, that is, the circuit is not in a stable state during these clock cycles after reset [22]. Additionally, in modern circuit designs containing clock gating the signals can start the propagation from the enabling clock edges after awaking during the functioning period, long after the reset. In this case the timing violation from the aggressive formulation may cause incorrect logic values and states, as pointed out in [26]. Furthermore, using the aggressive formulation we need to compute the converged arrival times $a_i^m$, which is not an easy task because of the coexistence of the maximum and minimum computations in (5) and (7). For example, the method in [27] relaxes the maximum and minimum computations by substituting them for the operators $\geq$ and $\leq$, respectively, and solves the resulting inequalities using linear programming, but this relaxation can not guarantee a solution for the unrelaxed problem as stated in [17], [26], [27].

Compared with the aggressive formulation, the conservative formulation can be verified very easily. For each latch $i$ we simply check the condition $a_i^1 \geq h_i$, where $h_i$ is a known random variable and $a_i^1 = \min_{j \to i}\{r_j + \lambda_{j,i}\}$ according to (5)–(7). This formulation is used in [26], [28] for static timing analysis and optimization. The simplified hold time verification with multiphase clocks explained in [23] is also a conservative formulation reasoned by assuming zero-delay shortest paths. In practice, hold time violations are mitigated by using multiphase clock schemes [14], [23], with which the yield of the circuit is barely affected by such constraints, as confirmed by the experimental results in Section IV. In this paper, we use the conservative formulation to extract hold time constraints. Details will be given in Section III-F.

### III. THE PROPOSED METHOD

In this section, we explain our method for statistical timing analysis of latch-controlled circuits. We first explain the modeling of clock signals with different clock periods $T$ in Section III-A. Thereafter we discuss the basic framework to compute the statistical minimum clock period in Section III-B. This framework includes two steps. In the first step, setup time constraints are handled using reduced iterations. This step will be elaborated in Section III-C. In the second step, constraints from nonpositive loops are extracted using graph transformations. These constraints have been defined and discussed in Section II-B and will be elaborated in Section III-D. Hold time constraints are handled in Section III-F.

In the following, all mentioned variables, except those which we clearly stated as constants, are random. The max and sum operations used in this paper are very general and do not take advantage of any properties of the underlying distributions of the random variables, so that the proposed method can easily be adapted to use any statistical timing engine [1]–[10].

### A. Clock Modeling

In statistical timing analysis the statistical minimum clock period is computed and then the yield at any given clock period can be evaluated using the CDF of the distribution. For edge-triggered circuits no special consideration needs to be taken for the clock signal because the change of the clock period $T$ simply means the proportional change of the time difference between successive clock edges. The yield has no relation to the duty cycle of the clock signal unless it is too small or too large to prevent the flip-flops from proper behavior. On the contrary, in latch-controlled circuits the arrival time propagation depends on the position of the enabling edge $r_i$ and the phase shift $E_{j,i}$ defined in Table I and illustrated in Fig. 2. We explain the general clock modeling used in this paper as follows.

For a general model which can handle different clock schemes, the clock phase shift $E_{j,i}$ and the enabling edge $r_i$ in the local time zone illustrated in Fig. 2 are represented as

$$E_{j,i} = \epsilon_{j,i}T + \upsilon_{j,i} \tag{15}$$
$$r_i = \epsilon_i T + \upsilon_i \tag{16}$$

where $\epsilon_{j,i}$ and $\epsilon_i$ are constants with $\epsilon_{j,i} \geq 0$ and $1 > \epsilon_i \geq 0$. $\upsilon_{j,i}$ and $\upsilon_i$ are random variables. If the clock phases are generated using a delay-based mechanism, $E_{j,i}$, and $r_i$ if necessary, can be assumed as known random variables. In this case, we set $\epsilon_{j,i} = 0$ and $\epsilon_i = 0$. If the clock phases are generated so that they scale proportionally, that is, $E_{j,i}$ and $r_i$ have fixed ratios to the clock period, both $\epsilon_{j,i}$ and $\epsilon_i$ are positive to model these ratios. If process variations in the clock circuitry are considered, the corresponding random variables can also be merged into $\upsilon_{j,i}$ and $\upsilon_i$, respectively.

### B. Basic Framework of the Proposed Method

In this section, we explain the basic idea of our method for computing the statistical minimum clock period. The details will be given in Section III-C and III-D.

In Section II-B, we have explained the algorithm published in [22] for static timing analysis of latch-controlled circuits with Algorithm 1. In this algorithm, the setup time constraints and loop constraints are verified separately. The arrival times are updated for $n$ clock cycles in L5–L10 of Algorithm 1. The setup time constraints must meet $A_i^n \leq T - s_i$ and the nonpositivity of the loops is checked using the condition $D_i^n \neq D_i^{n-1}$. But in statistical timing analysis the combina-tional delays are random variables usually in parametric form and therefore the iterations cost much more runtime than in the static case. Additionally the nonpositivity verification can not work anymore because random variables are described using moments and the correlation between them is implicitly contained in the parametric representations. Even though the moments of the departure times can be compared after the $n$ iterations, the coefficients representing correlation can rarely be matched.

**Algorithm 2**: Simplified constraint verification

| | |
|---|---|
| L1 | **foreach** *node i in RTG* **do** |
| L2 | $A_i^0 = r_i$ |
| L3 | **end** |
| L4 | **for** $m = 1$ **to** $n$ **do** |
| L5 | **foreach** *node i in RTG* **do** |
| L6 | $A_i^m = \max_{j \to i}\{A_j^{m-1} + \Lambda_{j,i}\}$ |
| L7 | **if** $A_i^m > T - s_i$ **then** |
| L8 | **return** *false* |
| L9 | **end** |
| L10 | **end** |
| L11 | **end** |
| L12 | **if** $no\_positive\_loops()$ *in RTG* **then** |
| L13 | **return** *true* |
| L14 | **else** |
| L15 | **return** *false* |
| L16 | **end** |

In the proposed method for statistical timing analysis, we use the same idea that the setup time constraints are extracted first. Thereafter we verify the nonpositivity of the loops using graph transformations. From both steps we establish a set of constraints for the clock period $T$, from which the statistical minimum can be computed. In this section, we use a simplified version of our algorithm to explain the basic idea and prove that it is equivalent to Algorithm 1. The complete algorithm will be given in Section III-C and III-D. This simplified algorithm is listed in Algorithm 2.

The arrival times in Algorithm 2 are iterated and setup times are verified directly in L4–L11. These iterations differ from the ones in Algorithm 1 in that the arrival times are updated without the comparison with the enabling clock edges and the positive loops are verified using graph transformations described by the function $no\_positive\_loops()$, whose details will be described in III-D. Using the following lemmas and theorem, we prove that Algorithm 2 and Algorithm 1 are equivalent so that the same set of timing constraints can be extracted.

**Lemma 1.** *All paths with $m$ edges and ending at node $i$ in the RTG are denoted by a path set $P_i^m$, $m \geq 1$. Assume the nodes on a path $p$ from $P_i^m$ are $j_0^p, j_1^p, \ldots j_m^p, j_m^p = i$. The arrival time $A_i^m$ in Algorithm 2 is equal to $\max_{p \in P_i^m}(r_{j_0^p} + \sum_{k=1}^m \Lambda_{j_{k-1}^p, j_k^p})$.*

*Proof:* Similar to the proof of the minimum distance in the Bellman-Ford algorithm [25], we prove this lemma using induction by $m$. If $m = 1$, $A_i^1$ is updated from all its fanin nodes with initial arrival times using L6 in Algorithm 2, so that the lemma is true. Assuming this lemma is correct for $m = l, l \geq 1$, for one of the fanin node $j$ of $i$ we have

$$A_j^l + \Lambda_{j,i} = \max_{p \in P_j^l}(r_{j_0^p} + \sum_{k=1}^{l} \Lambda_{j_{k-1}^p, j_k^p}) + \Lambda_{j,i}$$

$$= \max_{p \in P_{j,i}^{l+1}}(r_{j_0^p} + \sum_{k=1}^{l+1} \Lambda_{j_{k-1}^p, j_k^p}) \quad (17)$$

where the path set $P_{j,i}^{l+1}$ contains all paths with $l+1$ edges and ending at $i$ through $j$. According to the maximum computation in L6 of Algorithm 2 we have

$$A_i^{l+1} = \max_{j \to i} \{A_j^l + \Lambda_{j,i}\}. \quad (18)$$

Because any path reaching $i$ must go through one of its fanin nodes we can conclude that $A_i^{l+1}$ meets the lemma by combining (17) and (18). ∎

**Lemma 2.** *In Algorithm 1, $A_i^m$ is equal to the maximum of the arrival times from the paths to $i$ with edge number smaller or equal to $m$, $m \geq 1$, that is, $A_i^m = \max_{q=1,2,\ldots m} \{\max_{p \in P_i^q}(r_{j_0^p} + \sum_{k=1}^{q} \Lambda_{j_{k-1}^p, j_k^p})\}$.*

*Proof:* We prove with induction by $m$. When $m = 1$, each of the path to $i$ contains only one edge and $A_i^1$ meets the lemma according to L3, L7 and L8 in Algorithm 1. Assume this lemma is correct with $m = l, l \geq 1$. Replacing the departure time in L7 by the right side of L8 of Algorithm 1, we have $A_i^{l+1} = \max_{j \to i}\{A_j^l + \Lambda_{j,i}, r_j + \Lambda_{j,i}\}$. Similar to the proof of Lemma 1, $\max_{j \to i}\{A_j^l + \Lambda_{j,i}\}$ is the maximum of the arrival times computed from paths with numbers of edges between 2 and $l+1$ and ending at $i$ according to the assumption at $m = l$. $\max_{j \to i}\{r_j + \Lambda_{j,i}\}$ is the maximum arrival time from paths with one edge and ending at $i$. Therefore the lemma is proved with $m = l + 1$. ∎

**Theorem 1.** *Algorithm 2 and Algorithm 1 are equivalent.*

*Proof:* If there is at least one positive loop in the RTG, Algorithm 1 returns $false$ according to [22]. This is equivalent to Algorithm 2. If there is no positive loop, Algorithm 1 checks if the maximum arrival time of all paths to a node violates the setup constraint in L12, according to Lemma 2. In Algorithm 2, these arrival times are directly verified with the setup constraints in L7–L9. If any violation exists, both algorithms return $false$. If all setup time constraints are met, both algorithms return $true$. This proves the equivalence of the two algorithms. ∎

Algorithm 2 differs from Algorithm 1 in that we directly check the arrival times with setup time constraints so that we do not need to save the maximum of the arrival times from paths with different numbers of edges during the iterations. This simplification makes the computation of the statistical minimum clock period fast and will be discussed further in Section III-C, where the detailed algorithm to extract setup time constraints by extending Algorithm 2 is shown.

With the simplified iterations explained in Algorithm 2 we will show how to extract the minimum clock period from the setup time and loop constraints. According to Lemma 1 the verification in L7 of Algorithm 2 checks whether the arrival times across the paths meet setup time constraints. For one of the path $j_0^p, j_1^p, \ldots j_m^p$ the constraint is written as

$$r_{j_0^p} + \sum_{k=1}^{m} \Lambda_{j_{k-1}^p, j_k^p} \leq T - s_i. \quad (19)$$

Using (15), (16) and the definition of $\Lambda_{j,i}$ in Table I, the left side of (19) can be transformed into

$$(\epsilon_{j_0^p}T + \upsilon_{j_0^p}) + \sum_{k=1}^{m}(\Delta_{j_{k-1}^p, j_k^p} - \epsilon_{j_{k-1}^p, j_k^p}T - \upsilon_{j_{k-1}^p, j_k^p}) =$$

$$(\epsilon_{j_0^p} - \sum_{k=1}^{m}\epsilon_{j_{k-1}^p, j_k^p})T + (\upsilon_{j_0^p} + \sum_{k=1}^{m}(\Delta_{j_{k-1}^p, j_k^p} - \upsilon_{j_{k-1}^p, j_k^p})). \quad (20)$$

The coefficient of $T$, $\epsilon_{j_0^p} - \sum_{k=1}^{m}\epsilon_{j_{k-1}^p, j_k^p}$ in (20), can be merged into a constant and written as $c_p$. The second term $\upsilon_{j_0^p} + \sum_{k=1}^{m}(\Delta_{j_{k-1}^p, j_k^p} - \upsilon_{j_{k-1}^p, j_k^p})$ is the sum of random variables and can be computed into a single random variable using an SSTA engine, written as $d_p$. Then the constraint (19) can be transformed into

$$c_pT + d_p \leq T - s_i \Longleftrightarrow T_p = (d_p + s_i)/(1 - c_p) \leq T \quad (21)$$

where $1 - c_p > 0$ because $\epsilon_{j_0^p}$ in (20) is smaller than 1 and all $\epsilon_{j_{k-1}^p, j_k^p}$ are nonnegative as defined in (15) and (16). The constraint (21) defines a lower bound for the clock period $T$, denoted by $T_p$. This should be extracted at all latches during the iterations in Algorithm 2 and the maximum of all the lower bounds is computed as the minimum period of the clock signals to guarantee the proper behavior of the circuit. Although we describe the arrival times in the form of (20), the paths are not enumerated during the iterations. Instead, the sums of constants and random variables are computed from the arrival times of the fanin nodes in each iteration, as shown in L6 of Algorithm 2. This is much faster than path enumeration because of the large number of paths in the RTG. The details of updating the arrival times and extracting the lower bound $T_p$ at each node from the setup time constraint will be explained in Section III-C.

As discussed at the beginning of this section, the nonpositivity of loops can not be checked using $D_i^n \neq D_i^{n-1}$ because both variables are in the parametric form. In our method, we use graph transformations to capture these constraints without enumerating all the loops. Details of this algorithm will be explained in Section III-D. According to the definition of the nonpositive loop, the constraint from loop $j_0, j_1, \ldots j_{n_l}, j_0 = j_{n_l}$ is

$$\sum_{k=1}^{n_l} \Lambda_{j_{k-1}, j_k} \leq 0 \Longleftrightarrow \quad (22)$$

$$\sum_{k=1}^{n_l}(\Delta_{j_{k-1}, j_k} - \epsilon_{j_{k-1}, j_k}T - \upsilon_{j_{k-1}, j_k}) \leq 0. \quad (23)$$

Let $\sum_{k=1}^{n_l}(\Delta_{j_{k-1}, j_k} - \upsilon_{j_{k-1}, j_k}) = d_l$ and $\sum_{k=1}^{n_l}\epsilon_{j_{k-1}, j_k} = c_l$, (23) creates a lower bound $T_l$ for the clock period $T$ as

$$T_l = d_l/c_l \leq T. \quad (24)$$

The constraints (21) from all arrival times and (24) from all loops guarantee that all the setup time constraints are extracted. In the first $n$ clock cycles, the arrival times meet the setup time constraints of all latches because all the constraints are verified during the iterations in Algorithm 2, L4–L11. After the $n$ cycles, the arrival times must have traversed a loop because the largest loop contains no more than all the nodes in the RTG. With the nonpositive constraints, the arrival times after traversing loops do not increase so that the setup time constraints are guaranteed. This is also the basis for the Bellman-Ford algorithm to avoid unnecessary iterations in finding the shortest paths in a graph.

*C. Setup Time Constraint Extraction with Reduced Iterations*

In this section we explain our method to extract the minimum clock period from setup time constraints in detail. Unlike other previously published methods [17]–[19] we do not assume the clock period $T$ is given and compute the parametric minimum clock period from all the lower bounds described in (21). We maintain a random variable $T_{min}$ to represent this minimum clock period and each time we extract a constraint in the form of (21) we update $T_{min}$ by

$$T_{min} \leftarrow \max\{T_{min}, T_p\}. \quad (25)$$

After the iterations in Algorithm 2 terminate, all the lower bounds $T_p$ defined in (21) for $T$ are merged into $T_{min}$. The condition $T_{min} \leq T$ guarantees that the arrival times meet the setup time constraints in the first $n$ clock cycles. Note that $T_{min}$ increases gradually during the iterations until the minimum clock period from setup time constraints is computed.

The assumption that $T$ is unknown causes problems in the arrival time propagation. The first is that the arrival times can not simply be described by a single random variable but with a coefficient for $T$, as shown in (20) and simplified in (21). We solve this problem by propagating $c_p$ and $d_p$ in a pair $(c_p, d_p)$ for the arrival time, representing $c_p T + d_p$. When an arrival time is computed from the arrival times of the fanin nodes, the edge weights, $\Lambda_{j,i}$ defined in Table I, are added to the pairs representing the arrival times at the fanin nodes from the last iteration. Since $\Lambda_{j,i}$ can also be represented by a pair of constant and random variable according to (2) and (15), namely,

$$\Lambda_{j,i} = \Delta_{j,i} - E_{j,i} = (-\epsilon_{j,i})T + (\Delta_{j,i} - v_{j,i}) = c_e T + d_e \quad (26)$$

the new arrival time is computed by adding the coefficients and random variables, respectively.

Because the arrival times of the fanin nodes may be propagated across different paths leading to different coefficients of $T$, it is possible that the computed new pairs have different coefficients. During updating the arrival times in L6 of Algorithm 2, if a newly computed pair has the same coefficient as one of the pairs that are already computed, only the maximum of the random variables is computed with which the arrival time is updated. This computation does not increase the number of saved pairs. But if the newly computed pair does not match any of the already computed pairs by comparing the coefficients of $T$, the new pair is appended to the saved pair list and indexed by its coefficient. This operation increases the number of the saved pairs and then the runtime in the following iterations.

The nonzero coefficients of $T$ in the propagation of arrival times are the motivation that we use Algorithm 2 instead of Algorithm 1 to extract the setup time constraints. In Algorithm 1 we need to save the maximum arrival times in the past $m$ iterations. These arrival times go through paths with $1, 2, \ldots m$ edges according to Lemma 2 so that the coefficients of $T$ are different and the arrival time pairs can not be merged. This costs much runtime because the arrival times of all the $m$ iterations are propagated in further iterations, which may not be needed because the constraints are captured by propagating the arrival times only from the paths with $m$ edges, as proved in Theorem 1. Additionally, the duplicated constraints increase the number of maximum computations to update $T_{min}$ in (25), potentially affecting the accuracy of the timing analysis due to the approximation during the statistical maximum computation.

We use several approaches together to reduce the runtime of updating the arrival times. The first two directly compress the saved pairs according to their statistical dominance. The third reduces the number of iterations and identifies a much smaller graph compared with the original one for evaluating the nonpositivity of the loops so that the runtime of the proposed method can be reduced further.

In the first approach, we compare the pairs of the arrival time at latch $i$ with the enabling clock edge $r_i$ each time after the arrival time is updated. Assume the arrival time $A_i^m$ has a pair $(c_p, d_p)$ which meets

$$c_p T + d_p \leq r_i = \epsilon_i T + v_i \iff (d_p - v_i) \leq (\epsilon_i - c_p) T \quad (27)$$

we remove this pair from the saved pair list because the further propagation of this pair does not affect $T_{min}$. To prove this reduction, assume the arrival time pair is propagated to node $i$ across a path $j_0, j_1, \ldots j_m, j_m = i$. According to Lemma 1 and Algorithm 2 the pair $(c_p, d_p)$ should be propagated further to check the setup time constraints at any following node. Assume this arrival time pair is propagated across a path $j_0, j_1, \ldots j_m, \ldots j_{m+q}, q \geq 1$, to node $j_{m+q}$. The setup time constraint at node $j_{m+q}$ is

$$c_p T + d_p + \sum_{k=m+1}^{m+q} \Lambda_{j_{k-1}, j_k} \leq T - s_{j_{m+q}}. \quad (28)$$

Combining (27) and (28) we find that the constraint created from propagating $(c_p, d_p)$ to a node after $i$ is always dominated by the propagation from the enabling edge $r_i$ so that the removal of $(c_p, d_p)$ does not affect the minimum clock period.

Because different latches may have different enabling clock edges, $\epsilon_i - c_p$ in (27) is not always positive. As shown in (20) $c_p$ is computed by subtracting the positive coefficients in the clock phase shifts, $\epsilon_i - c_p$ nevertheless becomes positive very quickly during the iterations. If $\epsilon_i - c_p$ is positive, (27) can be transformed into

$$(d_p - v_i)/(\epsilon_i - c_p) \leq T. \quad (29)$$

Because $T$ is unknown, condition (29) can not be evaluated directly. According to (25) we know $T_{min} \leq T$, although $T_{min}$ may still not be large enough to capture all setup time constraints because not all iterations are finished. We then compare the left side of (29) with $T_{min}$ using

$$(d_p - v_i)/(\epsilon_i - c_p) \leq T_{min}. \quad (30)$$

If (30) is true the condition (29) is guaranteed. Both sides of (30) are random variables so that it can be true only with a certain probability, defined as
$$p_r = Prob\{(d_p - v_i)/(\epsilon_i - c_p) \leq T_{min}\}. \quad (31)$$
If $p_r$ approximates 1 we remove the pair from the pair list for the arrival time.

The compression above can effectively reduce the number of pairs representing the arrival times, because the edge delays across paths compensate each other. Additionally, $T_{min}$ increases gradually when new constraints are merged, thus strengthening the effect of the arrival time removal by applying (31). Moreover, we also extract the nonpositive loop constraints from self loops in the original RTG before running the iterations. This provides an initial value for $T_{min}$ in the comparison so that the probability $p_r$ is increased in the early iterations. The self loop extraction will be explained later in Section III-D.

The second approach to reduce the runtime compares the pairs of an arrival time with each other. These pairs have different coefficients of $T$ due to different clock phase shifts on the paths. For two pairs $(c_{p_1}, d_{p_1})$ and $(c_{p_2}, d_{p_2})$, the former is dominated by the latter and can be removed if
$$c_{p_1}T + d_{p_1} \leq c_{p_2}T + d_{p_2} \iff d_{p_1} - d_{p_2} \leq (c_{p_2} - c_{p_1})T. \quad (32)$$
In case $c_{p_2} - c_{p_1} > 0$, (32) is equivalent to
$$(d_{p_1} - d_{p_2})/(c_{p_2} - c_{p_1}) \leq T. \quad (33)$$
Similar to (29)–(31), we remove the pair $(c_{p_1}, d_{p_1})$ if the probability
$$Prob\{(d_{p_1} - d_{p_2})/(c_{p_2} - c_{p_1}) \leq T_{min}\} \quad (34)$$
approximates 1.

The third approach in our method to reduce runtime implements an event-driven updating mechanism. As described in (27)–(31), the computed arrival time from a fanin node is compared with the enabling edge and not propagated if the probability $p_r$ in (31) approximates 1. During the iterations, newly created lower bounds from the setup time constraints are updated into $T_{min}$ using (25), causing $T_{min}$ to increase gradually. Therefore, the probability $p_r$ becomes larger and many arrival times are blocked by the corresponding enabling edges. Based on this fact, an event-driven arrival time updating can be implemented to reduce the number of iterations. In [22], [26] a similar event-driven mechanism is applied, but our method can terminate the iterations very early and return a subgraph for loop evaluation.

We implement this event-driven mechanism by maintaining an index $\zeta_i$ for node $i$ in the RTG. Each time the arrival time of node $i$ is updated in L6 of Algorithm 2 we change $\zeta_i$ to the value of the iteration index $m$. If a node has all its fanin nodes not updated in the last iteration, this node is not processed in the current iteration and its index is not changed, reflecting the fact that no valid arrival times from the previous stages can reach the current node. Owing to the compression methods described before many nodes are not updated as the iterations continue and therefore have their indexes $\zeta_i$ smaller than the current iteration index $m$.

This event-driven mechanism does not miss setup time constraints that affect the minimum clock period $T_{min}$. Consider a path $j_0, j_1, \ldots j_k, \ldots j_m$. If the arrival time of node $j_m$ is not updated at the $m$th iteration because all the arrival times of its fanin nodes are not updated, we can trace back across this path from $j_m$ until a node $j_k$ is found whose arrival time from $j_0$ is dominated by the enabling edge $r_k$ and blocked as described in (27)–(31). Therefore the arrival time from $j_0$ to $j_m$ is dominated by the arrival time from $j_k$ to $j_m$. If the arrival time from $j_k$ to $j_m$ can reach $j_m$, the corresponding setup time constraint is directly merged into $T_{min}$ during the iterations; otherwise it must be blocked at an intermediate node between $j_k$ and $j_m$ and the back tracing can be performed again until $j_{m-1}$ is reached. In this case the timing constraint is extracted in the first iteration and surely covered by $T_{min}$.

The complete algorithm extracting setup time constraints using the arrival time compression and event-driven iterations is listed in Algorithm 3, which is an extended version of Algorithm 2. By L6–L8 only the nodes with fanin nodes that are updated in the last iteration are visited, implementing the event-driven mechanism. The function $compress\_arrival\_time(A_j^m, r_j)$ compares the arrival time pairs computed for $A_j^m$ with the enabling clock edge using (27)–(31) and with each other using (32)–(34) to remove the dominated pairs. If the arrival time of a node has a nonempty pair list after being processed in the current iteration, the update index $\zeta_i$ of the node is changed to the current iteration index $m$ in L13 so that these nodes are marked as recently updated. If all arrival time pairs are removed, say, blocked by the enabling edge, this node is considered not updated and its update index is not changed. The function $update\_T_{min}(A_j^m, s_j)$ extracts the lower bound $T_p$ for $T$ using (19)–(21) and updates $T_{min}$ with (25).

The event-driven mechanism not only reduces the number of nodes to be processed in the iterations but also provides an early quit condition denoted by the function $quit\_cond()$ in Algorithm 3. In real circuits, the transparent propagation

---

**Algorithm 3**: Setup time constraint extraction

L1 **foreach** *node $i$ in RTG* **do**
L2     $A_i^0 = r_i$
L3     $\zeta_i = 0$
L4 **end**
L5 **for** $m = 1$ **to** $n$ **do**
L6     **foreach** *node $i$ with $\zeta_i = m-1$* **do**
L7         **foreach** *fanout node $j$ of $i$* **do**
L8             **if** $A_j^m$ *is not computed* **then**
L9                 $A_j^m = \max_{t \to j}\{A_t^{m-1} + \Lambda_{t,j}\}$
L10                 $compress\_arrival\_time(A_j^m, r_j)$
L11                 **if** $A_j^m$ *has a nonempty pair list* **then**
L12                     $update\_T_{min}(A_j^m, s_j)$
L13                     $\zeta_j = m$
L14                 **end**
L15             **end**
L16         **end**
L17     **end**
L18     **if** $quit\_cond()$ **then**
L19         **return** $RTG_r$
L20     **end**
L21 **end**
L22 **return** *RTG*

usually does not pass many latch stages, and even if some critical paths can continue the arrival time propagation, most of the paths are blocked very quickly during the iterations. Based on this observation the function $quit\_cond()$ identifies the critical part of the circuit to reduce the number of iterations.

Assume the current iteration index is $m$ and the number of nodes whose arrival times are updated in one of the iterations from $m - \tau$ to $m$, $m > \tau \geq 0$, is $n_r$. The subgraph composed of these $n_r$ nodes and all the edges between them is denoted by $RTG_r$. If $\tau \geq n_r$, the function $quit\_cond()$ returns true so that Algorithm 3 quits and returns $RTG_r$, whose loops will be checked for nonpositivity as will be described in Section III-D. The loop constraints from $RTG_r$ and the setup time constraints extracted by Algorithm 3 can sufficiently define the minimum clock period $T_{min}$. We prove this conclusion using the following lemma and theorem.

**Lemma 3.** *If $quit\_cond()$ returns true at the mth iteration with $m > \tau \geq 0$ and $\tau \geq n_r$, any simple loop in the RTG with a node not from $RTG_r$ is nonpositive under the condition $T_{min} \leq T$.*

*Proof:* We construct a path with $m$ nodes $j_0, j_1, \ldots j_{m-\tau}, \ldots j_m$ by traversing a loop containing a node not from $RTG_r$ as many times as necessary. If the loop contains more than $m$ nodes, the constructed path is only a part of the path forming the loop. From $j_{m-\tau}$ to $j_m$ we have $\tau + 1$ nodes and at least one of them is not updated in the $m - \tau$ to $m$ iterations because the subgraph $RTG_r$ contains only $n_r$ nodes with $n_r \leq \tau$ and we only consider simple loops where no embedded loops are involved. We denote this node not from $RTG_r$ as $j_{q_r}$. Because $m \geq q_r \geq m - \tau$ the arrival time of $j_{q_r}$ is not updated at the $q_r$th iteration so that we can find a node $j_{q_0}$ between $j_1$ and $j_{q_r}$ and the arrival time from $j_0$ to $j_{q_0}$ is dominated by the enabling edge $r_{j_{q_0}}$. Otherwise, if the arrival times from $j_0$ to all the nodes $j_1, j_2, \ldots j_{q_r}$ are not dominated by the corresponding enabling edges, the arrival time at $j_{q_r}$ must be updated by the propagation across the path $j_0, j_1, \ldots j_{q_r}$ or by an arrival time from another path that dominates the one across $j_0, j_1, \ldots j_{q_r}$, as explained for the function $compress\_arrival\_time(A_j^m, r_j)$ in Algorithm 3.

Then we repeat the process above to find a node $j_{q_1}$ after $j_{q_0}$ by constructing a path from $j_{q_0}$ across the loop so that the arrival time propagated starting from $j_{q_0}$ is dominated by the enabling edge $r_{j_{q_1}}$. We repeat this process $l$ times and denote the cumulative weight of the path from node $j_{q_k}$ to node $j_{q_{k+1}}$ by $\Lambda_{j_{q_k} \to j_{q_{k+1}}}$ so that we have

$$r_{j_{q_k}} + \Lambda_{j_{q_k} \to j_{q_{k+1}}} \leq r_{j_{q_{k+1}}}, \quad 0 \leq k < l. \quad (35)$$

Adding both sides of (35) for all $k$ from 0 to $l-1$, we have

$$\sum_{k=0}^{l-1} r_{j_{q_k}} + \sum_{k=0}^{l-1} \Lambda_{j_{q_k} \to j_{q_{k+1}}} \leq \sum_{k=0}^{l-1} r_{j_{q_{k+1}}} \iff \quad (36)$$

$$r_{j_{q_0}} + \sum_{k=0}^{l-1} \Lambda_{j_{q_k} \to j_{q_{k+1}}} \leq r_{j_{q_l}} \iff \quad (37)$$

$$r_{j_{q_0}} + \Lambda_{j_{q_0} \to j_{q_l}} \leq r_{j_{q_l}}. \quad (38)$$

If $l$ is large enough, the path from $j_{q_0}$ to $j_{q_l}$ traverses the loop many times. Therefore, the cumulative weight of the loop must be nonpositive to guarantee that (38) is correct for any $l$ and the finite $r_{j_{q_0}}$ and $r_{j_{q_l}}$. Otherwise, the left side of (38) may exceed any $r_{j_{q_l}}$ by traversing the loop enough times. ∎

**Theorem 2.** *In Algorithm 3 if in the $m - \tau$ to $m$ iterations the number of updated nodes is no larger than $\tau$, that is, $n_r \leq \tau$, and the subgraph $RTG_r$ composed of these $n_r$ nodes and the edges between them has no positive loop, the setup time constraints across any path in the RTG, even containing loops, are sufficiently met.*

*Proof:* This proof is similar to the proof for Lemma 3. According to Lemma 3 and the assumption that $RTG_r$ contains no positive loop, we know all the loops in the RTG are nonpositive, so that we only need to prove the theorem for the paths without loops. Consider the path $j_0, j_1, \ldots j_{n_p}$ with $n_p$ edges. If $n_p \leq m$, the setup time constraints are already extracted in the $m$ iterations of Algorithm 3 before the function $quit\_cond()$ returns true. If $n_p > m$, the path can be renumbered to $j_0, j_1, \ldots j_{m-\tau}, \ldots j_m, \ldots j_{n_p}$. Similar to the proof of Lemma 3, we can find a node $j_{q_0}$ between the nodes $j_{m-\tau}$ and $j_m$ so that the arrival time from $j_0$ is dominated by the enabling edge $r_{j_{q_0}}$. We then move the starting node to $j_{q_0}$ and can find a similar node between $j_{q_0+m-\tau}$ and $j_{q_0+m}$ and repeat this process until the node $j_{n_p}$ is covered. Because the distance between each starting node and the next one is no larger than $m$, the setup time constraints are guaranteed by the first $m$ iterations which propagate arrival times from the enabling edges of all the nodes at the same time, as shown in Algorithm 3. Because we do not limit $n_p$ the lemma is proved for any path. ∎

A special case of Lemma 3 and Theorem 2 is that $\tau = 0$. Then Algorithm 3 only checks whether no node is updated in the latest iteration. This exit condition is used in [29, Algorithm 4] to terminate the iterations and bypass the checking of loop nonpositivity because in this case the number of nodes in $RTG_r$ is zero. Using this exit condition can avoid to verify the nonpositivity of the loops in $RTG_r$ in some test cases. But usually more iterations are needed to reach this exit condition and therefore the algorithm in [29] consumes more runtime than the method in this paper. This special case of exit condition is listed as the following corollary.

**Corollary 1.** *With the condition $T_{min} \leq T$ all the loops in the RTG are nonpositive if no arrival time in an iteration is updated. In this case all the timing constraints are extracted by Algorithm 3.*

In Lemma 3 and Theorem 2 $\tau$ can have any integer value if $m > \tau \geq 0$. In the implementation of $quit\_cond()$ we check the condition starting from $\tau = 0$ and the function returns true anytime when the condition $n_r \leq \tau$ is met. Because the node number $n_r$ is smaller than $\tau$, a smaller $\tau$ leads to a smaller $RTG_r$ for the nonpositivity checking, therefore reducing the runtime in the following step.

Algorithm 3 returns a graph and all the loops in this graph must be nonpositive. As discussed in Section III-B, these constraints can not be verified using the arrival time comparison, L12 in Algorithm 1. In the next section, we propose a method to extract these timing constraints using graph transformations.

*D. Nonpositive Loop Constraint Extraction*

According to Lemma 3 and Theorem 2 all the loops in $RTG_r$ must be checked for nonpositivity and the constraints of (22)–(24) should be extracted and merged into $T_{min}$ using (25). Verifying nonpositive loop constraints is already studied in [17]–[19]. These methods use iterative or loop-breaking algorithms to traverse loops and need much runtime. In this section we propose a method to solve this problem using graph transformations. The proposed method can also be used to analyze the yield of a latch-controlled circuit with the constraint graph in [17].

To extract timing constraints from the loops, we propose three graph transformations: serial merge, parallel merge, and self loop removal. The first two transformations reduce the number of nodes and edges in the graph and create self loops representing original loops in the graph. The third transformation removes the self loops from the graph and extracts timing constraints using (22)–(25).

**Definition 3.** *For a node $j_k$ in the graph, the **serial merge** removes $j_k$ and all edges connected to it. Between each pair of fanin node $j_{k-1}$ and fanout node $j_{k+1}$, a direct edge is created with weight $\Lambda_{j_{k-1},j_k} + \Lambda_{j_k,j_{k+1}}$. This transformation is denoted by function $serial\_merge(j_k)$.*

If a node without a self loop is removed using the serial merge transformation, the cumulative weight of any loop in the graph does not change. Assume a loop with $n_l$ nodes, $j_0, j_1, \ldots j_{k-1}, j_k, j_{k+1}, \ldots j_{n_l}$ with $j_0 = j_{n_l}$, $n_l \geq 2$. After $j_k$ is removed in the transformation, a new edge is created between $j_{k-1}$ and $j_{k+1}$ with weight $\Lambda_{j_{k-1},j_k} + \Lambda_{j_k,j_{k+1}}$. Therefore the cumulative weight of the loop is unchanged. The serial merge transformation is also used in [12] to reduce the timing graph for statistical time analysis and in [30] to create timing models for combinational circuits.

In Fig. 4(a) an example of the serial merge is illustrated, where node 4 in Fig. 1 is removed. The new edges which are created between the fanin and fanout nodes of the removed node 4 are marked by ∥. Because node 1 is both the fanin and the fanout node of 4, a self loop is created at node 1 representing the original loop $1 \to 4 \to 1$.

After applying the serial merge, edges sharing the same starting and ending nodes may appear. For example, an edge from 1 to 3 is created in Fig. 4(a). In the original graph in Fig. 1, there is already an edge from 1 to 3. These edges are called *parallel edges* and merged into one edge using the following transformation.

**Definition 4.** *In the RTG, if two edges share the same starting node $j_k$ and ending node $j_{k+1}$, they can be merged into one edge with the weight equal to the maximum of the weights of the removed edges. The new edge weight is computed by statistical maximum if the coefficients of the original edge weights are equal, or is determined by the edge weight which has statistical dominance to the other one. This transformation is called **parallel merge** and denoted by function $parallel\_merge(j_k, j_{k+1})$.*

The parallel merge does not affect any nonpositive loop constraints. Assume two edges from node $j_k$ to $j_{k+1}$ exist and there is a path from $j_{k+1}$ to $j_k$. This path and either of the parallel edges form a loop. The nonpositive constraint requires that the cumulative weights of both loops, and therefore their maximum, should be no larger than 0. Because both loops share all the path except the edge from $j_k$ to $j_{k+1}$, the maximum cumulative weight can be computed by the cumulative weight from $j_{k+1}$ to $j_k$ plus the maximum of the weights of the two edges from $j_k$ to $j_{k+1}$. Therefore the parallel merge does not affect the nonpositive loop constraints. The parallel merge transformation is also used in [12] to compress the timing graph.

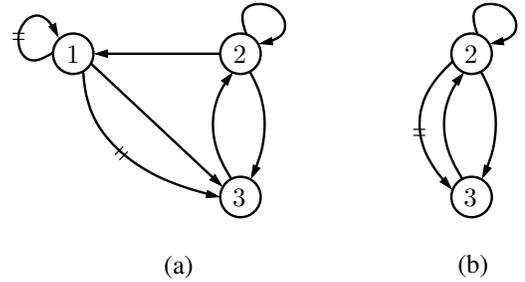

Fig. 4. Serial merge transformation example

As explained in Section II an edge weight is $\Lambda_{j,i}$ equal to $(-\epsilon_{j,i})T + (\Delta_{j,i} - v_{j,i})$ according to (2) and (15), and can be represented by a constant-variable pair $(c_e, d_e)$. In the serial merge, the sum of two edges can be computed by adding the constants and the random variables, respectively. But in the parallel merge, the maximum can only be performed directly when the constants in the two edge weights are equal. In this case, the constant remains unchanged and the maximum of the two random variables is computed using an SSTA engine. This limitation may cause many parallel edges not to be merged and the number of edges in the graph to increase very fast. We solve this problem using the statistical dominance concept similar to the processing of arrival times shown in (32)–(34). If an edge is dominated by another parallel edge, the former is removed from the graph to reduce the number of edges. The loop constraints from the graph are not affected by this edge removal because the loops across the dominating edge remain in the graph for loop constraint extraction.

In the serial merge, self loops are created representing the loops through different nodes in the original graph, for example, the self loop at node 1 in Fig. 4(a). These loops are processed by the third graph transformation defined below.

**Definition 5.** *The self loops from the original graph or the ones created by the serial merge transformations are removed from the graph by the transformation called **self loop removal** and the corresponding loop constraints are extracted using (22)–(24). The lower bounds for the clock period extracted from the self loops are merged into $T_{min}$ using (25). This transformation is denoted by the function $remove\_self\_loop(j)$, where $j$ is the node having self loops.*

The proposed graph transformations can be applied repeatedly until there is no node in the graph. Because the serial merge and the parallel merge do not affect the cumulative weights, all the loop constraints can be captured by the self loops during the transformations. This process is illustrated further in Fig. 4(b), where node 1 in Fig. 4(a) is removed by applying the proposed graph transformations. Similarly one of the nodes in Fig. 4(b) can be removed using the serial merge transformation with only one node left in the graph thereafter and all loop constraints are captured from the self loops.

**Algorithm 4**: Loop constraint extraction

| | |
|---|---|
| L1 | **foreach** node $i$ in the graph **do** |
| L2 | $\quad remove\_self\_loop(i)$ |
| L3 | **end** |
| L4 | **while** a node exists in the graph **do** |
| L5 | $\quad i = next\_node()$ |
| L6 | $\quad serial\_merge(i)$ |
| L7 | $\quad$ **foreach** node $j$ in the fanin nodes of $i$ **do** |
| L8 | $\quad\quad remove\_self\_loop(j)$ |
| L9 | $\quad$ **end** |
| L10 | $\quad$ **if** parallel edges exist between $j_k$ and $j_{k+1}$ **then** |
| L11 | $\quad\quad parallel\_merge(j_k, j_{k+1})$ |
| L12 | $\quad$ **end** |
| L13 | **end** |

In the example of Fig. 4(a), the number of edges in the graph does not increase after the transformation. In practice many nodes have a large number of fanin and fanout nodes. The removal of these nodes may result in a large number of edges in the graph and therefore a proper order of nodes for the serial transformations is required. Although theoretically an optimum order exists to remove the nodes with a minimum number of edges after each transformation, this order is not easy to determine in a complex graph. In our method we use a heuristic order to remove the nodes in view of the fact that Algorithm 3 returns only a small part of the graph with dominating loops for the transformations. For a node $i$ with $n_{in}^i$ input edges and $n_{out}^i$ output edges, the number of the new edges created after the serial merge is $\eta_i = n_{in}^i \times n_{out}^i$. We first remove the nodes with smaller $\eta_i$, implying a relatively small increase of the number of edges. The newly created edges may be merged to other parallel edges or directly removed if self loops are formed. The selection of the node with the minimum $\eta_i$ is denoted by the function $next\_node()$. The loop constraint extraction using graph transformations is listed in Algorithm 4.

*E. Discussions on Extracting Setup Time Constraints*

In the proposed method, setup time constraints are extracted by running Algorithm 3 first. This algorithm returns a subgraph containing the dominating loops for applying the graph transformations using Algorithm 4. In Algorithm 4 L1–L3 are used to remove all the self loops in the original graph, so that in L7–L9 only the nodes with edges to $i$ are checked for newly formed self loops. The removal of self loops in the original graph, L1–L3 of Algorithm 4, can also be run before Algorithm 3 so that $T_{min}$ has an initial value to be used to compress the arrival times more effectively using (27)–(34).

In Section III-A we described a clock modeling which can handle fixed-ratio clock scaling and delay-based clock scaling. In the first case the constants $\epsilon_{j,i}$ and $\epsilon_i$ in (15) and (16) are positive and the corresponding timing constraints can be extracted using Algorithm 3 and 4 directly. In the second case we set $\epsilon_{j,i}$ in (15), and $\epsilon_i$ in (16) if necessary, to zero so that the phase shift and the enabling clock edge are modeled using the random variables $v_{j,i}$ and $v_i$, respectively. Consequently, the edge weights $\Lambda_{j,i}$ defined in (2) become purely random variables. In this case Algorithm 3 and 4 can still be applied to extract the timing constraints with small revisions explained as follows.

As all edge weights are known random variables, the arrival time described on the left side of (19) is also known since the coefficients of $T$ in $r_{j_0^p}$ and $\Lambda_{j_{k-1}^p,j_k^p}$ are all zero. Therefore the arrival time propagation described in Section III-C to III-D needs not to use the pair $(c_p, d_p)$ since all $c_p$ are zero. The setup time constraints in this case still define lower bounds for the clock period $T$. Like the arrival time propagation in Section III-C these lower bounds can be merged into $T_{min}$ during the iterations using (25). In the comparison of the arrival times with enabling clock edges in (27) or with each other in (32) the coefficients $\epsilon_i$, $c_p$, $c_{p_1}$, and $c_{p_2}$ are all zero so that the comparison probability defined and used in (31) and (34) can be computed directly, making the runtime of the iterations even shorter.

After the iterations finish, the graph returned by Algorithm 3 should be processed by Algorithm 4. For each loop the left side of the nonpositive constraint (22) is a known random variable so that (22) does not define a lower bound for the clock period $T$. Instead, all the loop constraints extracted by Algorithm 4 are merged together as

$$C_l = \max_{l \in L}\{\sum_l \Lambda_{j_{k-1},j_k}\} \leq 0 \qquad (39)$$

where the sum computes the cumulative weight of the loop $l$ belonging to the loop set $L$, which are captured by Algorithm 4. To guarantee the proper behavior of the circuit, the constraints (25) and (39) must be met. For a given clock period $T_g$ the yield of the circuit is computed by

$$Prob\{T_{min} \leq T_g \wedge C_l \leq 0\} =$$
$$Prob\{\max\{T_{min} - T_g, C_l\} \leq 0\} \qquad (40)$$

where $\wedge$ means *logic and*. The maximum in (40) can be computed using an SSTA engine so that the yield at $T_g$ can be evaluated.

*F. Hold Time Constraints*

As discussed in Section II-C we use the conservative formulation to extract the hold time constraints. According to (5)–(8) the constraint for a pair of latches with an edge from $j$ to $i$ is written as

$$a_i^1 \geq h_i \iff r_j + \delta_{j,i} - E_{j,i} \geq h_i. \qquad (41)$$

In the case that the phase shift and the rising clock edge are generated using delay-based mechanism discussed in Section III-A, the constraints from all the latch pairs can be merged as

$$C_h = \max_{j,i}(h_i - r_j - \delta_{j,i} + E_{j,i}) \leq 0 \qquad (42)$$

where $C_h$ is a known random variable and the maximum computation is performed for all pairs of the nodes with an edge between them. For the second case that the clock ratios are fixed as modeled by (15) and (16), we have the constraints

as
$$(\epsilon_j T + v_j) + \delta_{j,i} - (\epsilon_{j,i} T + v_{j,i}) \geq h_i \iff \quad (43)$$
$$(\epsilon_j - \epsilon_{j,i}) T \geq h_i - \delta_{j,i} - v_j + v_{j,i}. \quad (44)$$

In the case $\epsilon_j - \epsilon_{j,i} > 0$ the constraint defines a lower bound $T \geq (h_i - \delta_{j,i} - v_j + v_{j,i})/(\epsilon_j - \epsilon_{j,i})$, which can be merged into $T_{min}$ using (25). If $\epsilon_j - \epsilon_{j,i} = 0$, the constraint can be merged into $C_h$ similar to (42). If $\epsilon_j - \epsilon_{j,i} < 0$, the constraint (44) defines an upper bound for the clock period. We denote this upper bound using $T_{max}$ and update it each time we have a new constraint as

$$T_{max} \leftarrow \min\{T_{max}, (h_i - \delta_{j,i} - v_j + v_{j,i})/(\epsilon_j - \epsilon_{j,i})\}. \quad (45)$$

Similar to (40) the yield of the circuit at a given clock period $T_g$ considering hold time constraints is computed as

$$Prob\{T_{min} \leq T_g \wedge C_h \leq 0 \wedge T_g \leq T_{max}\} =$$
$$Prob\{\max\{T_{min} - T_g, C_h, T_g - T_{max}\} \leq 0\}. \quad (46)$$

## IV. EXPERIMENTAL RESULTS

The proposed method was implemented in C++ with a single thread and tested using a 2.67 GHz CPU. The cells in the benchmark circuits were mapped to a 90 nm library from an industry partner. The standard deviations of transistor length, oxide thickness and threshold voltage were assigned to 15.7%, 5.3% and 4.4% of the nominal values, respectively [31]. The cell delays were created using the method proposed in [1]. We used the SSTA engines proposed in [2] and [32] to compute the sum and maximum of random variables.

Mainly two cases have been tested in our experiments. In the first case, the ISCAS89 benchmark circuits were used and all registers were assumed as latches with a single clock so that the runtimes can be compared with the test cases in other previously published methods. In the second case, we constructed the circuits from the ISCAS89 benchmarks using the method in [26], as illustrated in Fig. 5 which is adapted from [26, Fig. 6]. The original circuit can be considered as a set of latches $L_1$ and the combinational logic $C_1$. We duplicated the latches and the combinational logic as $L_2$ and $C_2$ and cross connected them with $L_1$ and $C_1$. These constructed circuits are referred by adding the prefix '2' to the names of the original ISCAS89 benchmarks. The clocks $clk\_1$ and $clk\_2$ formed a two-phase clock scheme and were set to inverted clocks with clock phase shift $T/2$ and duty cycle equal to 0.5 during the test of setup time constraints.

To verify the accuracy of the proposed method, we ran Monte Carlo simulation with 10 000 samples for each benchmark circuit. With each sample, the method proposed in [14] was used to compute the minimum clock period. The Monte Carlo simulation was implemented with parallelism due to the runtimes for large circuits which exceed 36 days for the simulation with a single-threading implementation. The runtimes of all the instances of the parallel Monte Carlo simulation were summed and the runtime of the code for parallelism was subtracted, so that the runtimes reported in this section can be considered as measured from a single-threading implementation and can be compared with the runtimes from the proposed method.

We first show the results of extracting setup time constraints using Algorithm 3 and 4 for case A and B in Table II

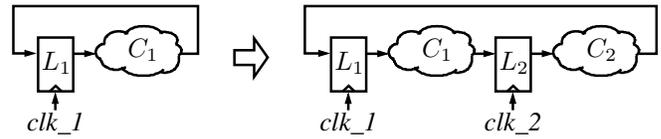

Fig. 5. Circuit construction for Case B

and III, respectively. In these tables $n_c$ and $n_s$ denote the numbers of combinational cells and sequential cells in the circuit, respectively. $\mathcal{E}_\mu$ is the relative error of the mean of the minimum clock period $T_{min}$ and defined as $(|\mu_{SSTA} - \mu_{MC}|/\mu_{MC}) \times 100$, that is, in percentage, where $\mu_{SSTA}$ and $\mu_{MC}$ are the means of the minimum clock period computed by the proposed method and by Monte Carlo simulation using the original circuit, respectively. Similar to $\mathcal{E}_\mu$, $\mathcal{E}_\sigma$ is defined to show the accuracy of the standard deviation of the minimum clock period. From $\mathcal{E}_\mu$ and $\mathcal{E}_\sigma$ we can see that the results of the proposed method are accurate, with the largest difference in the results as 1.51%. We have also computed the clock periods at which the yields are 97% from the results of the proposed method and Monte Carlo simulations. The average errors of case A and B are 0.39% and 0.35%, respectively. The similar yield error in [17] is 0.24%, which is 1.03% according to the re-implementation of this method in [18]. The average error of the method proposed in [18] is 0.21%. The maximum error of the proposed method in this paper is slightly larger than the ones in [17], [18], but less than 1%, still smaller than the maximum error of the re-implementation of [17] in [18]. By this comparison, we can further confirm the accuracy of the proposed method.

The main advantage of the proposed method compared with other existing methods is its efficiency. The runtimes of the proposed method and Monte Carlo simulation are shown as $t_P$ and $t_{MC}$ in seconds in Table II and III, respectively, where the runtimes with 4 decimal places are measured by running the proposed method 100 times and computing the average, because these runtimes are too small to be directly measured with accuracy. Comparing the speedup ratios of the proposed method to Monte Carlo simulation with the ones in other existing methods, for example, [17]–[19], the proposed method has more than 10 times speed advantage while still maintaining good accuracy.

The high computational efficiency is achieved mainly by reducing the number of the iterations drastically using the exit condition $quit\_cond()$ in Algorithm 3. The numbers of the iterations that are actually needed in capturing the setup time constraints are listed in Table II and III and denoted by $m$.

TABLE II
RESULTS OF SETUP TIME CONSTRAINT EXTRACTION FROM CASE A

| Circuit | $n_c$ | $n_s$ | $\mathcal{E}_\mu(\%)$ | $\mathcal{E}_\sigma(\%)$ | $m$ | $n_r$ | $t_P$ (s) | $t_{MC}$ (s) | Speedup |
|---|---|---|---|---|---|---|---|---|---|
| s298 | 119 | 14 | 0.06 | 0.45 | 3 | 0 | 0.0024 | 9.39 | 3913 |
| s526 | 193 | 21 | 0.52 | 0.42 | 3 | 0 | 0.0041 | 15.04 | 3668 |
| s820 | 289 | 5 | 0.13 | 0.42 | 2 | 0 | 0.0045 | 6.95 | 1544 |
| s1238 | 508 | 18 | 0.24 | 0.53 | 2 | 0 | 0.0061 | 9.27 | 1520 |
| s1423 | 657 | 74 | 0.10 | 0.16 | 23 | 22 | 0.05 | 425.60 | 8512 |
| s5378 | 2779 | 179 | 0.76 | 0.49 | 41 | 37 | 0.12 | 1869.44 | 15579 |
| s9234 | 5597 | 211 | 0.40 | 0.17 | 31 | 28 | 0.34 | 1666.35 | 4901 |
| s13207 | 7951 | 638 | 0.69 | 0.86 | 3 | 2 | 0.47 | 2496.50 | 5312 |
| s15850 | 9772 | 534 | 1.30 | 1.51 | 6 | 0 | 0.96 | 9926.95 | 10341 |
| s38584 | 19253 | 1426 | 0.96 | 1.12 | 7 | 0 | 1.69 | 20916.45 | 12377 |

TABLE III
RESULTS OF SETUP TIME CONSTRAINT EXTRACTION FROM CASE B

| Circuit | $n_c$ | $n_s$ | $\mathcal{E}_\mu(\%)$ | $\mathcal{E}_\sigma(\%)$ | $m$ | $n_r$ | $t_P(s)$ | $t_{MC}(s)$ | Speedup |
|---|---|---|---|---|---|---|---|---|---|
| 2s298   | 238   | 28   | 0.42 | 0.62 | 19  | 16  | 0.0007 | 53.98      | 77114  |
| 2s526   | 386   | 42   | 0.21 | 0.94 | 32  | 28  | 0.0167 | 153.98     | 9220   |
| 2s820   | 578   | 10   | 0.01 | 0.00 | 10  | 10  | 0.0119 | 32.75      | 2752   |
| 2s1238  | 1016  | 36   | 0.05 | 1.03 | 2   | 0   | 0.02   | 41.80      | 2090   |
| 2s1423  | 1314  | 148  | 0.09 | 0.49 | 127 | 126 | 1.02   | 12582.80   | 12336  |
| 2s5378  | 5558  | 358  | 0.72 | 0.49 | 132 | 124 | 0.55   | 8421.37    | 15312  |
| 2s9234  | 11194 | 422  | 1.41 | 0.83 | 117 | 112 | 1.37   | 40624.57   | 29653  |
| 2s13207 | 15902 | 1276 | 0.91 | 0.39 | 40  | 22  | 1.19   | 91808.01   | 77150  |
| 2s15850 | 19544 | 1068 | 0.12 | 0.61 | 264 | 258 | 13.31  | 1194933.90 | 89777  |
| 2s38584 | 38506 | 2852 | 0.06 | 1.29 | 256 | 254 | 26.42  | 3181881.83 | 120435 |

TABLE IV
RESULTS OF SETUP AND HOLD TIME CONSTRAINT EXTRACTION

| Circuit | $\mathcal{E}_{y,1}(\%)$ | $\mathcal{E}_{y,2}(\%)$ | $\mathcal{E}_{y,3}(\%)$ | $t_P(s)$ | $t_{MC}(s)$ | Speedup |
|---|---|---|---|---|---|---|
| 2s298   | 1.21 | 1.30 | 0.13 | 0.10  | 46.85      | 469    |
| 2s526   | 0.66 | 0.05 | 0.01 | 0.17  | 136.44     | 803    |
| 2s820   | 0.33 | 0.81 | 0.05 | 0.21  | 34.26      | 171    |
| 2s1238  | 1.90 | 0.18 | 0.20 | 0.28  | 47.04      | 174    |
| 2s1423  | 0.25 | 0.50 | 1.11 | 1.30  | 12879.18   | 10386  |
| 2s5378  | 2.52 | 1.57 | 0.26 | 1.30  | 9475.90    | 8312   |
| 2s9234  | 0.20 | 1.05 | 0.55 | 2.58  | 42094.30   | 20045  |
| 2s13207 | 1.64 | 0.86 | 0.70 | 2.69  | 88006.82   | 45599  |
| 2s15850 | 0.61 | 1.46 | 2.29 | 14.68 | 1268328.19 | 96819  |
| 2s38584 | 0.26 | 0.76 | 0.71 | 27.56 | 3309694.24 | 138423 |

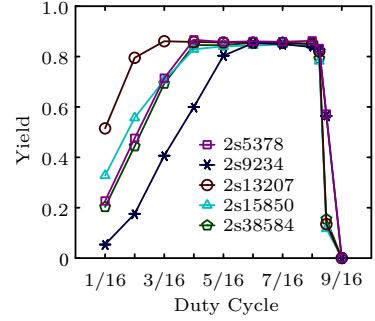

Fig. 6. Duty cycle and yield

These numbers are rather small compared with the number of latches, $n_s$ in Table II and III, especially for the large circuits for which the runtime cost of one iteration is also large. After applying Algorithm 3 the returned graph for checking nonpositivity using graph transformations contains $n_r$ nodes so that only a very small portion of the original RTG needs to be processed using the graph transformations, therefore reducing the runtime further.

In addition to Case A and B, we have tested larger circuits constructed by two rows and two columns of the circuits in Fig. 5. The connections between the modules are generated randomly. The largest circuit in this setting contains more than 150k cells. Compared to the runtimes in Table III, the runtimes for these circuits are 10 times larger on average. This nonlinear scalability is due to the multiple-level transparency from all latches and inherent in the timing analysis of latch-controlled circuits [14], [17], [18]. The proposed method, however, consumes no more than 200 seconds for the large circuits in this test setting, which is still much less than the runtimes of other existing methods tested using smaller circuits. We have run Monte Carlo simulations for the circuits constructed from 2s298 to 2s5378. The simulations for other circuits require several months so that are unaffordable even with parallelism. The maximum relative error of mean and standard deviation from the accuracy comparison is 1.58%, which is still acceptable for statistical timing analysis.

To incorporate the hold time constraints into the analysis we changed the duty cycle of the two clocks in Case B to 0.515 so that there is some overlap between the active periods of the clock signals. With this setting hold time violations will cause the yield to decrease so that the accuracy of the proposed method can be tested. We first ran Monte Carlo simulations with this setting for all the circuits in Case B. Then we selected three clock periods, $T_1 = \mu_{MC}$, $T_2 = \mu_{MC} + \sigma_{MC}$ and $T_3 = \mu_{MC} + 2 \times \sigma_{MC}$ for each circuit to compute the yields using the proposed method, where $\mu_{MC}$ and $\sigma_{MC}$ are the mean and standard deviation of the minimum clock period from Monte Carlo simulation, respectively. The errors of the yields from the proposed method to the ones from Monte Carlo simulation are shown as $\mathcal{E}_{y,1}$, $\mathcal{E}_{y,2}$ and $\mathcal{E}_{y,3}$ in Table IV, respectively, where the runtimes are also shown as $t_P$ and $t_{MC}$. The comparison confirms that the proposed method can predict the yield very well as the hold time constraints are considered within a very short runtime.

In latch-controlled circuits the length of the active periods of the clock signals affects the transparency and correspondingly the yield. The longer the active period is, the more chance the data signal can be propagated transparently. To evaluate this relation, we changed the clock duty cycle in Case B from 1/16 to 9/16 and measured the yields at the clock $T_2$ defined before. The results are illustrated in Fig. 6. The yields of the circuits increase with the larger duty cycles, but do not change after the duty cycle reaches 6/16, where the clock period is mainly determined by the loop nonpositivity. In Fig. 6 the yields start to decrease when the duty cycle is near to 0.5, from which the clock periods have overlap so that the hold time constraints have more effect on the yields. The yields at duty cycle 9/16 are all zero because there are always hold time violations in this setting. This trend confirms that in multiphase clock schemes the clocks are normally designed without overlap between the active periods to reduce the hold time constraint violation [14], [23]. Because no yield increase after the duty cycle reaches 6/16, unnecessary long active periods of the clocks should be avoided in clock designs. In this case even the hold time constraints need not to be verified because they only affect the yields when overlap between the active periods of the clock signals is about to happen. The case we have discussed is based on the conservative hold time formulation described in Section II-C. The conclusion here also applies to the aggressive formulation where the earliest arrival times used for verifying hold time constraints are larger than the ones in the conservative formulation and the hold time constraints may already be met when the duty cycles are set to keep the clocks from overlap.

## V. CONCLUSION

In this paper, we proposed a method for statistical timing analysis of latch-controlled circuits. This method extracts setup time constraints and evaluates the nonpositivity of loops sep-arately. The former is processed by a reduced iterative propagation and the latter is handled using graph transformations. The resulting minimum clock period is in a parametric form from which the yield of the circuit at any given clock period can be computed. Hold time constraints are also incorporated. Experimental results confirm that the proposed method is much faster than other previously published methods while maintaining good accuracy for the minimum clock periods as well as the yields.

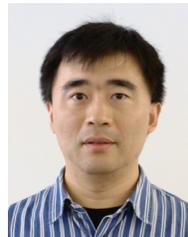

**Bing Li** received the Bachelor and Master degrees in communication and information engineering from Beijing University of Posts and Telecommunications in 2000 and 2003, respectively, and the PhD degree from Technische Universität München (TUM), Munich, Germany, in 2010. Currently he is doing postdoctoral research in the Institute for Electronic Design Automation, TUM. His research interests include timing/power analysis and circuit optimization.

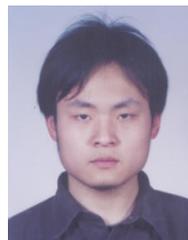

**Ning Chen** received the Dipl.-Ing. degree in electrical engineering and information technology from Technische Universität München (TUM), Munich, Germany, in 2007. Since 2008, he has been a Research and Teaching Assistant with the Institute for Electronic Design Automation, TUM. His current research interests include electronic design automation for digital circuits with focus on timing analysis.

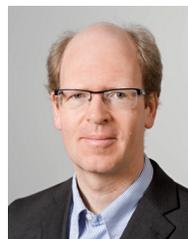

**Ulf Schlichtmann** (S'88–M'90) received the Dipl.-Ing. and Dr.-Ing. degrees in electrical engineering and information technology from Technische Universität München (TUM), Munich, Germany, in 1990 and 1995, respectively. From 1994 to 2003, he was with Siemens AG and Infineon Technologies AG, where he held various technical and management positions in design automation, design libraries, IP reuse, and product development. Since 2003, he has been with TUM as Professor and Head of the Institute for Electronic Design Automation. His research interests are in computer-aided design of electronic circuits and systems, with special emphasis on designing robust systems. From 2008 to 2011, he served as Dean of TUM's Department of Electrical Engineering and Information Technology.